\documentclass[twocolumn,trackchanges]{aastex6}
\usepackage{color,amsmath,textcomp,url,graphicx,subfigure,latexsym}
\usepackage{epsfig}
\usepackage{latexsym}%
\usepackage{graphicx}%
\usepackage{amssymb}%
\usepackage{url}
\usepackage{afterpage}
\usepackage{ltablex}
\usepackage{threeparttable}

\newcommand{\chapternote}[1]{{%
  \let\thempfn\relax
  \footnotetext[0]{\emph{#1}}
}}

\newcolumntype{C}[1]{>{\centering\let\newline\\\arraybackslash\hspace{0pt}}m{#1}}

\newcommand{\kms}{\hbox{km\,s$^{-1}$}}

\newcommand{\MJup}{$M_{\mathrm{Jup}}$}

\newcommand{\Teff}{$\ensuremath{T_{\mathrm{eff}}}$}

\newcommand{\target}{WISE J0711--5736}

\shorttitle{}
\shortauthors{Kellogg et al.}

\begin{document}

\title{Discovery of a Possible Early-T Thick-Disk Subdwarf from the AllWISE2 Motion Survey$^*$}
\author{Kendra Kellogg\altaffilmark{1,2}, J. Davy Kirkpatrick\altaffilmark{2}, Stanimir Metchev\altaffilmark{1,3}, Jonathan Gagn\'e\altaffilmark{4,5}, Jacqueline K. Faherty\altaffilmark{6}}
\altaffiltext{1}{Department of Physics and Astronomy, Centre for Planetary and Space Exploration, The University of Western Ontario, 1151 Richmond St, London, ON N6A 3K7, Canada; kkellogg@uwo.ca}
\altaffiltext{2}{Infrared Processing and Analysis Center, Mail Code 100-22, California Institute of Technology, 1200 E. California Blvd., Pasadena, CA 91125, USA}
\altaffiltext{3}{Department of Physics and Astronomy, Stony Brook University, Stony Brook, NY, 11794-3800, USA}
\altaffiltext{4}{Carnegie Institution of Washington DTM, 5241 Broad Branch Road NW, Washington, DC~20015, USA}
\altaffiltext{5}{NASA Sagan Fellow}
\altaffiltext{6}{Department of Astrophysics, American Museum of Natural History, Central Park West at 79th Street, New York, NY 10034, USA}

\begin{abstract}
We have discovered a potential T0 $\pm$ 1 subdwarf from a search for sources in the AllWISE2 Motion Survey that do not have counterparts in surveys at shorter wavelengths. With a tangential velocity of $\sim$170~\kms, this object --- WISE J071121.36--573634.2 --- has kinematics that are consistent with the thick-disk population of the Milky Way. Spectral fits suggest a low-metallicity for this object but also allow for the possibility of unresolved multiplicity. If \target\ is indeed an sdT0 dwarf, it would be only the second early-T subdwarf discovered to date.

\end{abstract}


\maketitle

\section{Introduction}

\chapternote{$^*$This paper includes data gathered with the 6.5 meter Magellan Telescopes located at Las Campanas Observatory, Chile.}

Brown dwarfs are the lowest mass products of star formation. With masses that are below the hydrogen burning minimum mass (HBMM; $<$75\MJup), they cannot undergo sustained nucleosynthesis like their higher mass brethren. As such, their compositions stay essentially unchanged from when they formed (e.g. \citealp{burrows93,marley96}). Metal-poor brown dwarfs and very low-mass stars offer unique insights into the young Milky Way as they are objects that formed in the early galaxy and their compositions have not changed much since that time.

Recently, dedicated searches for the old, low-mass population of objects in our galaxy have been carried out using the Wide-field Infrared Survey Explorer (WISE; \citealp{pinfield14,kirkpatrick14,kirkpatrick16}), UKIDSS Large-Area Survey (ULAS; \citealp{zhang17a}), Two-Micron All Sky Survey (2MASS; \citealp{kirkpatrick10}), and Sloan Digital Sky Survey (SDSS; \citealp{sivarani09}). The totality of the results of these surveys reveals that the lowest-mass stars and brown dwarfs with low metallicity are relatively rare. The very lowest metallicity low-mass stars and brown dwarfs belong to the galactic halo of the Milky Way and are the oldest objects in the galaxy. At a slightly younger age and higher metallicity are the members of the thick-disk. Both populations have high space velocities and are very uncommon in the solar neighborhood. Only $\sim$6\% and $<$1\% of the brown dwarfs in the vicinity of the sun are expected to be part of either the thick-disk or halo populations, respectively \citep{gazzano13}. These brown dwarfs have primordial compositions and are excellent testing grounds for models at low metallicities. 

L/T transition subdwarfs, in particular, are of key interest in studying the effects of metallicity on cloud physics in atmospheric models. The transition between L and T dwarfs, in the case of solar-metallicity brown dwarfs, is a rapid evolution from cloudy to clear atmospheres (e.g.\ \citealp{ackerman01,burrows01,burgasser02b,marley02}) and takes place over a narrow effective temperature range of only 200--300K (e.g.\ \citealp{kirkpatrick05}). Late-L and early-T subdwarfs, then, are excellent objects to study the role that metallicity plays in the evolution of brown dwarfs and their atmospheres. Due to the nature of the L/T transition, however, this spectral range is comparatively devoid of objects, even more so than the late-T subdwarfs which are not readily detected due to their intrinsically faint near-infrared magnitudes (only four late-T subdwarfs known; discussed in $\S$\ref{sec:halo}).

There has also been evidence that there is a ``subdwarf gap" which divides stellar and sub-stellar objects at late ages on a color-magnitude diagram (e.g.\ \citealp{kirkpatrick16,zhang17b}). Stellar theory predicts that at $\sim$10~Gyr, the lowest mass stars and brown dwarfs diverge in their observable properties due to the different energy production process above and below the hydrogen-burning minimum mass (e.g.\ \citealp{chabrier97,burrows01,burgasser04c}). Low-mass stars and brown dwarfs are more difficult to distinguish at younger ages but evolution models predict the stellar/substellar boundary occurs at higher temperatures for solar metallicity objects ($\sim$L2.5; \citealp{dieterich14}) and brown dwarfs have spectral types as early as M6 at very young ages (e.g.\ \citealp{burrows01,baraffe15,dupuy17}). Searches for low-metallicity objects reveal that the subdwarf gap appears to be between mid- and late-L dwarfs \citep{kirkpatrick16,zhang17b}. Late-L/early-T subdwarfs, then, would be on the substellar side of this region. The exact extension of the gap, however, has yet to be determined and locating the edges is of keen interest in the study of low-metallicity objects.

Given the rarity of halo/thick-disk brown dwarfs, finding even one new object to study is a significant step forward in understanding this population. An even more important step to understanding old, low-metallicity objects and their atmospheres is filling in the gap between late-L and late-T subdwarfs. As part of the AllWISE2 Motion Survey (hereafter AllWISE2; \citealp{kirkpatrick16}), we have possibly identified the second object in this spectral range that also has kinematics consistent with the population of thick-disk/halo objects --- WISE J071121.36--573634.2. 

In $\S$\ref{sec:selection} we discuss how we identified this object and in $\S$\ref{sec:obs} we outline our observations and data reduction. We discuss the spectral classification of \target\ in $\S$\ref{sec:class} and put it in context of the population of halo/thick-disk objects in $\S$\ref{sec:halo}.

\section{Candidate Selection} \label{sec:selection}
The AllWISE1 and AllWISE2 Motion Surveys \citep{kirkpatrick14,kirkpatrick16} were searches designed to leverage the multi-epoch mid-infrared observations of the entire sky and identify objects with proper motions detectable over the 1-year period of the AllWISE mission. Objects with such high apparent motions are either located in the solar neighborhood or have inherently large space velocities. The AllWISE1 Motion Survey imposed an {\it rchi2/rchi2\_pm} $>$ 1.03, selecting objects where the $\chi^2$ value of the stationary fit was at least 3\% higher than the $\chi^2$ of the motion fit. This criteria was removed for the AllWISE2 Motion Survey allowing objects with smaller motions to potentially be recovered in the second iteration of the survey. An interesting subset of the motion sources are ones that do not have counterparts in surveys at shorter wavelengths (e.g SDSS and 2MASS) --- typically late-T and Y dwarfs. The model predictions for the spectral energy distributions of such objects peak at $\sim$4-12\micron\ (700K$>\Teff>$250K) so they are very often undetected in optical and near-IR surveys. We, therefore, created a devoted search for similar objects.

We published 58 WISE-only candidates from AllWISE1 in \cite{kirkpatrick14}. We implemented the same type of selection criterion for WISE-only sources in AllWISE2, selecting objects that did not have counterparts in 2MASS or SDSS. After visual verifications (details in section 2 of \citealp{kirkpatrick16}), we ended up with 11 WISE-only objects from AllWISE2, for a total of 69 candidate late-T and Y dwarfs from both AllWISE1 and AllWISE2. The WISE magnitudes and proper motions of the 11 new WISE-only AllWISE2 candidates are presented in Table~\ref{tab:cand}. One AllWISE1 object turned out to be an important new discovery: the Y dwarf WISEA J085510.74--071442.5 \citep{luhman14a}. Four of the candidates from AllWISE2 were also already known T and Y dwarfs: WISE J223617.60+510551.8 (T5; \citealp{mace13a}), WISE J104752.35+212417.2 (T6.5; \citealp{burgasser99}), WISE J115013.85+630241.5 (T8; \citealp{kirkpatrick11}), and WISE J140518.39+553421.3 (Y0pec?; \citealp{kirkpatrick11}). Two of the known ultra-cool dwarfs --- WISE J104752.35+212417.2 and WISE J223617.60+510551.8 --- actually did have 2MASS counterparts but their motions are high enough that they were not recognized in the 2MASS images. For the rest of the analysis, we only consider the 64 new candidate ultra-cool dwarfs.

\begin{deluxetable*}{ccccccccc}
\tabletypesize{\scriptsize}
\tablecolumns{9}
\tablewidth{0pt}
\tablecaption{AllWISE2 Motion Candidates Lacking 2MASS Counterparts
\label{tab:cand}}
\tablehead{
\colhead{Designation} & \colhead{W1} & \colhead{W2} & \colhead{AllWISE} & \colhead{AllWISE} & \colhead{Calculated} & \colhead{Calculated} & \colhead{Spectral} & \colhead{Discovery}  \\
\colhead{$ $} & \colhead{(mag)} & \colhead{(mag)} & \colhead{(R.A. Motion)} & \colhead{(Decl. Motion)} & \colhead{$\mu_{\alpha}$cos($\delta$)} & \colhead{$\mu_{\delta}$} & \colhead{Type} & \colhead{Publication}  \\
\colhead{$ $} & \colhead{$ $} & \colhead{$ $} & \colhead{(mas\,yr$^{-1}$)} & \colhead{(mas\,yr$^{-1}$)} & \colhead{(mas\,yr$^{-1}$)} & \colhead{(mas\,yr$^{-1}$)} & \colhead{$ $} & \colhead{$ $} }
\startdata
\object{WISE J003428.12+393153.7} & 15.910 $\pm$ 0.046 & 14.666 $\pm$ 0.050 & 663 $\pm$ 224 & $-$1033 $\pm$ 233 & $-$3 $\pm$ 25 & $-$48 $\pm$ 80 & & 1 \\
\object{WISE J071121.36$-$573634.2} & 15.092 $\pm$ 0.029 & 14.627 $\pm$ 0.038 & 507 $\pm$ 137 & 998 $\pm$ 145 & 18 $\pm$ 10 & 990 $\pm$ 90 & & 1 \\
\object{WISE J081031.30$-$475602.8} & 15.741 $\pm$ 0.039 & 14.398 $\pm$ 0.041 & $-$1215 $\pm$ 272 & $-$902 $\pm$ 274 & 6 $\pm$ 25 & 63 $\pm$ 80 & & 1 \\
\object{WISE J082811.56$-$443738.1}\tablenotemark{a} & 14.751 $\pm$ 0.031 & 12.650 $\pm$ 0.024 & 512 $\pm$ 83 & 643 $\pm$ 88 & 2 $\pm$ 28 & 18 $\pm$ 85 & & 1 \\
\object{WISE J102055.17+530859.4} & 15.666 $\pm$ 0.042 & 14.526 $\pm$ 0.047 & 1247 $\pm$ 304 & 1287 $\pm$ 326 & $-$2 $\pm$ 25 & 31 $\pm$ 80 & & 1 \\
\object{WISE J104752.35+212417.2} & 15.377 $\pm$ 0.036 & 13.004 $\pm$ 0.030 & $-$908 $\pm$ 171 & $-$682 $\pm$ 186 & $-$112 $\pm$ 25 & $-$516 $\pm$ 80 & T6.5 & 2 \\
\object{WISE J115013.85+630241.5} & 16.958 $\pm$ 0.089 & 13.405 $\pm$ 0.028 & 330 $\pm$ 198 & $-$1194 $\pm$ 202 & 63 $\pm$ 25 & $-$540 $\pm$ 80 & T8 & 3 \\
\object{WISE J122738.12$-$232819.6} & 16.102 $\pm$ 0.058 & 14.692 $\pm$ 0.061 & 1897 $\pm$ 317 & 221 $\pm$ 352 & 7 $\pm$ 27 & 33 $\pm$ 85 & & 1 \\
\object{WISE J140518.39+553421.3} & 18.765 $\pm$ 0.396 & 14.097 $\pm$ 0.037 & $-$1862 $\pm$ 326 & $-$324 $\pm$ 334 & $-$265 $\pm$ 25 & 187 $\pm$ 80 & Y0(pec?) & 3 \\
\object{WISE J153747.73+181151.3} & 14.198 $\pm$ 0.025 & 12.220 $\pm$ 0.022 & $-$228 $\pm$ 64 & $-$367 $\pm$ 69 & 5 $\pm$ 25 & 48 $\pm$ 80 & & 1 \\
\object{WISE J223617.60+510551.8} & 13.827 $\pm$ 0.025 & 12.499 $\pm$ 0.025 & 573 $\pm$ 99 & $-$118 $\pm$ 89 & 70 $\pm$ 27 & 358 $\pm$ 85 & T5 & 4 \\
\enddata
\tablecomments{Discovery papers are: $^1$This work, $^2$\cite{burgasser99}, $^3$\cite{kirkpatrick11}, $^4$\cite{mace13a}.} 
\tablenotetext{a}{\scriptsize WISE J082811.56--443738.1 is a flux transient rather than a motion object.}
\end{deluxetable*}

All of the new objects had motions that were quite small and at the limit of bye-eye detection. Since the longest baseline for any of the objects was only 0.5 yr, we turned to the new data from the NEOWISE Reactivation mission \citep{mainzer11} to provide a longer baseline. We used the WISE co-add tool\footnote{\url{http://irsa.ipac.caltech.edu/applications/ICORE/}} developed at IPAC to combine images for each of the 64 new candidates for comparison. We created two co-added images: one from the observations during the first observational epoch (early 2010; WISE All-Sky) and one from observations during the last epoch (late 2015; NEOWISE-R). This gave a baseline of $>$5 years for most candidates, sufficient to authenticate their motion.  We determined the motions by measuring the centroids of the objects in each co-added image with DAOFIND in IRAF and calculating a linear fit between the objects at each of the two epochs. Because the epochs are $>$5 years apart for most candidates, the space motion of the objects dominates the observed motion and the parallax is a small, second-order component. Calculated proper motions based on the longer baseline co-added images are presented in Table~\ref{tab:cand}. One object, WISE J082811.56--443738.1, turned out to increase in brightness rather than have any motion so we did not follow this object up. 

\begin{figure*}
\centering 
\includegraphics[width=\textwidth]{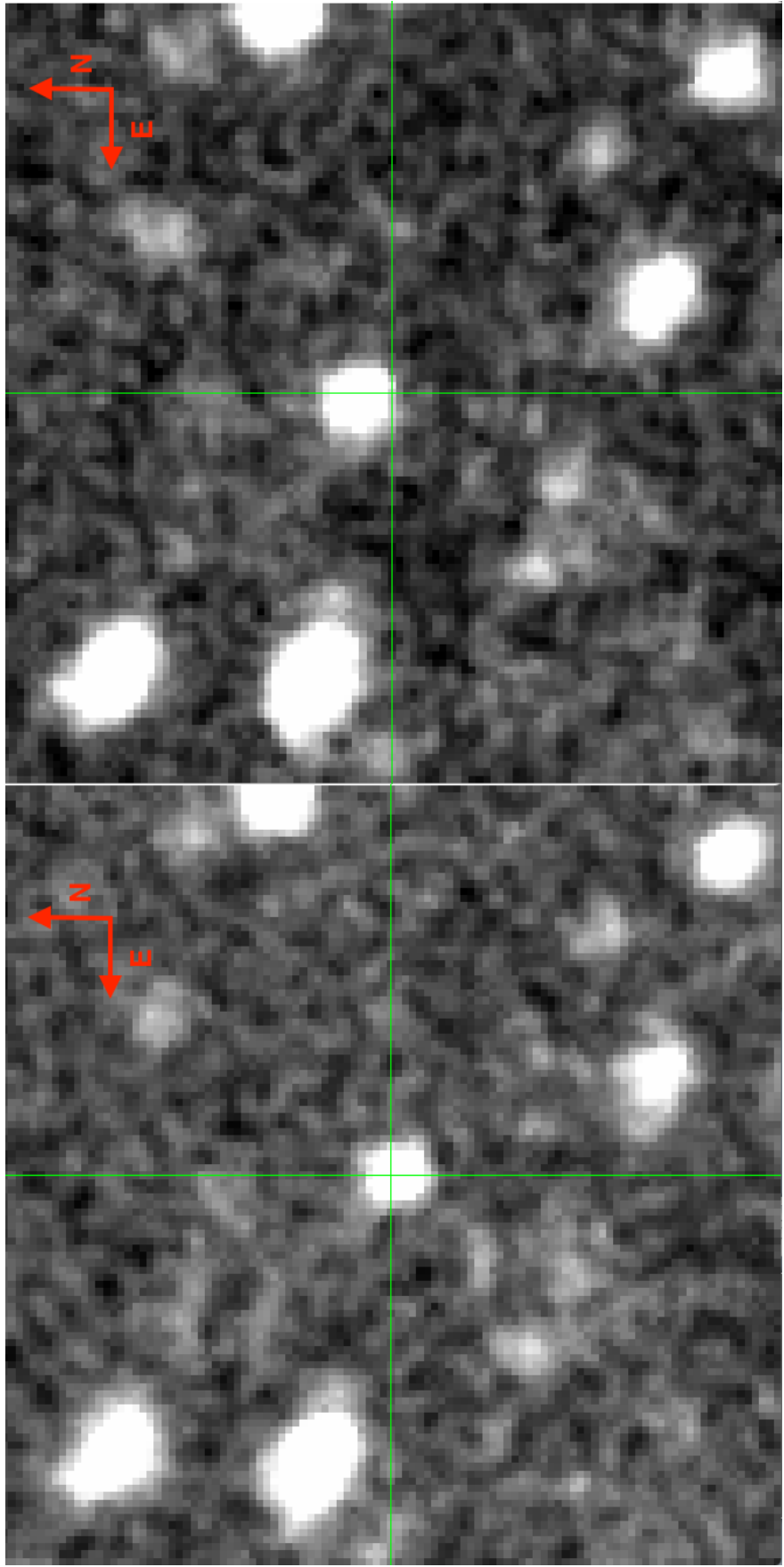}
\caption{\footnotesize Co-added 2'$\times$2' images from WISE in W2. The first image is from UT 2010 May 4 (WISE All-Sky), the second is from UT 2015 Nov 16 (NEOWISE-R).}
\label{fig:img}
\end{figure*}

Only one new candidate showed clear motion --- WISE J071121.36--573634.2 (hereafter \target). The calculated motions of all of the other new candidates were much smaller than the motions from AllWISE and are likely not brown dwarfs in the solar neighborhood. The co-added W2 images of \target\ from the first and last observational epoch are shown in Figure~\ref{fig:img}. From these images, we calculated a more accurate proper motion of \target\ (Table \ref{tab:target}) compared to the AllWISE proper motion estimates (Table~\ref{tab:cand}).

This object, however, does not have the characteristic infrared colors of a late-T or Y dwarf. Instead, the colors appeared to be that of a late-L or early-T dwarf based on the W1--W2 vs spectral type relation from \cite{kirkpatrick11}. With this assumption and an estimated absolute $W2$ magnitude for this spectral range from \cite{faherty16}, we calculated a photometric distance (36--39 pc) and tangential velocity (165--200~\kms) consistent with the population of galactic halo/thick-disk objects. We also calculated the reduced proper motion (H$_{W2} = W2$ $+$ 5log($\mu$) $+$ 5) of this object to be $\approx$19.6 mag. From Figure 11 of \cite{pinfield14}, we see that the H$_{W2}$ and $W1-W2$ values put this object far outside the 100~\kms\ curve --- also consistent with being part of at least the thick-disk population. For verification of the spectral type of \target\, we obtained near-IR spectra. The details are outlined in the next section.

\begin{tabularx}\textwidth{cc}
\caption{Object Properties
\label{tab:target}} \\
\hline \hline \\[-2ex]

\endhead

\hline \\[-2ex]
\multicolumn{2}{l}{$^1$Calculated motions are based on the positions on UT 2010} \\
\multicolumn{2}{l}{May 4 (WISE All-Sky) and 2015 Nov 16 (NEOWISE-R).} \\
\endfoot

Identifier & \object{WISE J071121.36--573634.2} \\
RA & 07:11:21.36 \\
Dec & $-$57:36:34.20 \\
W1 & 15.092 $\pm$ 0.029 mag\\
W2 & 14.627 $\pm$ 0.038 mag\\
W3 & 12.504 $\pm$ 0.500 mag\\
W4 & 9.558 $\pm$ 0.500 mag\\
Calculated $\mu_{\alpha}$cos($\delta$)$^1$ & 18 $\pm$ 10 mas\,yr$^{-1}$ \\
Calculated $\mu_{\delta}$$^1$ & 990 $\pm$ 90 mas\,yr$^{-1}$ \\
H$_{W2}$ & 19.61 $\pm$ 0.02 mag \\ [0.5ex]
\end{tabularx}

\section{Spectroscopic Observations and Data Reduction} \label{sec:obs}
We observed \target\ on UT 2016 December 14 with the Folded-port InfraRed Echellette (FIRE; \citealp{simcoe08,simcoe13}) at the \emph{Magellan}/Baade telescope in the high-throughput prism mode. The weather was clear with a seeing of $\sim$ 0\farcs3. We used the 0\farcs6 slit, yielding a resolving power of $R \sim 450$ in the 0.8--2.45\,$\mu$m wavelength range. We obtained four 120\,s exposures in a nodding ABBA pattern along the slit at airmasses of 1.262--1.275, yielding a signal-to-noise ratio of $\sim$ 50 per pixel in the 1.5--1.8\,$\mu$m range. The telluric exposures which were obtained immediately before \target\ were saturated and therefore not usable. The A0-type star HD 35265 observed on UT 2016 January 22 with six 1\,s exposures under clear weather conditions, a seeing of $\sim$ 0\farcs6, and an airmass of 1.210--1.231, was therefore used to perform the telluric correction. We obtained eleven 1\,s exposures of NeAr calibration lamps at the beginning of the respective nights to perform the wavelength calibration, as well as eleven exposures of low- and high-voltage dome flat fields to correct for pixel response variation. 

We reduced the data with the Interactive Data Language (IDL) Firehose v2.0 package (\citealp{bochanski09,gagne15d}\footnote{Avaliable at \url{https://github.com/jgagneastro/FireHose\_v2/tree/v2.0}}; see \citealp{gagne15b} for more details on the reduction package). We extracted the spectra using an optimal extraction approach. The local background was modeled using a basis spline fit to the masked aperture profile and subtracted it from the spectra. We subsequently extracted the spectra using a weighted profile extraction approach. We wavelength-calibrated and median-combined the resulting spectra using a modified version of the Spextool routine xcombspec \citep{cushing04}.  Finally, we corrected for telluric absorption and flux-calibrated the spectra with xtellcor\_general.pro using the A0 calibration star. 

\begin{figure}[t!]
\centering
\includegraphics[width=1\linewidth]{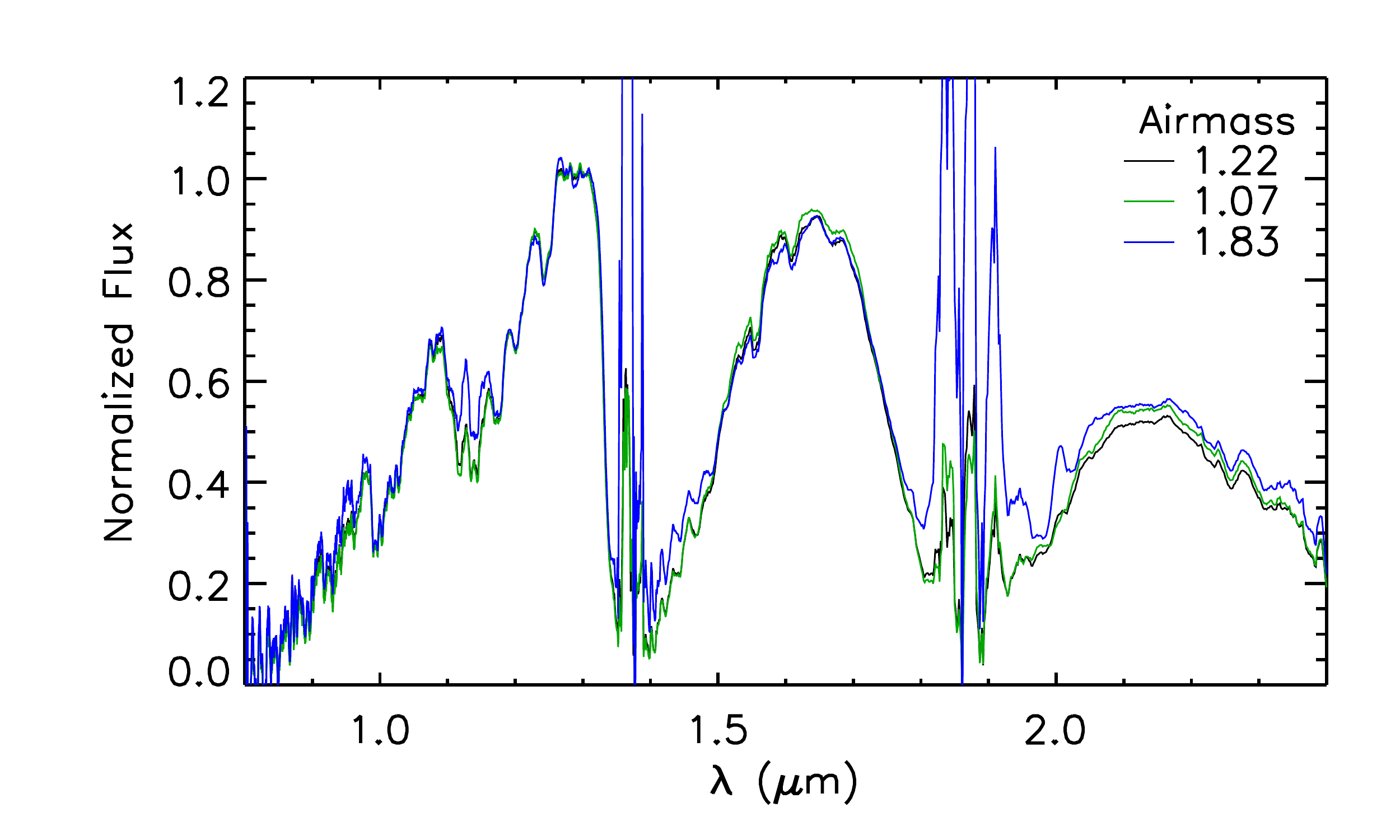}
\caption{\footnotesize Comparison of smoothed FIRE prism spectra of \target\ telluric corrected with HD 35265 (airmass $\approx$ 1.22; black), HD 102338 (airmass $\approx$ 1.07; green) and HD 149818 (airmass $\approx$ 1.83; blue). Spectra are normalized to the average flux in the 1.25--1.30\micron\ region.}
\label{fig:tellcomp}
\end{figure}

Since our telluric standard was observed 11 months prior to \target\ and the precipitable water content in the atmosphere was likely not the same on the two nights, we performed the same type of telluric correction using two additional standards. Observed at an airmass of $\sim$1.07 on 2016 January 22, HD 102338 was located in a region of sky with lower water content than HD 35265 (airmass $\approx$ 1.22) and HD 149818, observed at an airmass of $\sim$1.83 on 2016 January 22, was located in a region of sky with higher water content. Figure~\ref{fig:tellcomp} shows the comparison of \target\ corrected with all three telluric standards and smoothed for clarity. As we can see, the depths of all the water bands (1.10--1.18\micron, 1.30--1.45\micron, and 1.70--1.95\micron) are similar between the spectra corrected with two standards at low and moderate airmasses (HD 102338 and HD 35265). The spectrum that was corrected with the standard at a higher airmass (HD 149818) has less water absorption. However, the residual telluric features at 2.0--2.1\micron\ indicate that the spectrum corrected with HD 149818 was over-corrected and too much water absorption was removed. We also note that the (sub)stellar water bands are wider than the telluric ones due to atmospheric pressures, so the wings are not affected by the telluric correction, as Figure~\ref{fig:tellcomp} shows.

From this analysis we confirm that performing a correction with HD 35265 was sufficient to remove all of the telluric absorption. Thus, we conclude that the depths of the water bands are intrinsic to \target\ and are not a consequence of correcting with a telluric standard that was observed on a night with different precipitable water content.

\section{Spectral Classification and Kinematics} \label{sec:class}

In order to classify \target, we compare our FIRE spectrum to spectra of published objects and to theoretical spectra. When comparing to published spectra, we smooth our medium-resolution spectrum to the resolution of the comparison object using a least-squares quadratic interpolation. All spectra are normalized to the median flux value in the 1.25--1.30\micron\ region.

\subsection{Late-L/Early-T Subdwarf} \label{subsec:type}

\begin{figure}[t!]
\centering
\includegraphics[width=1\linewidth]{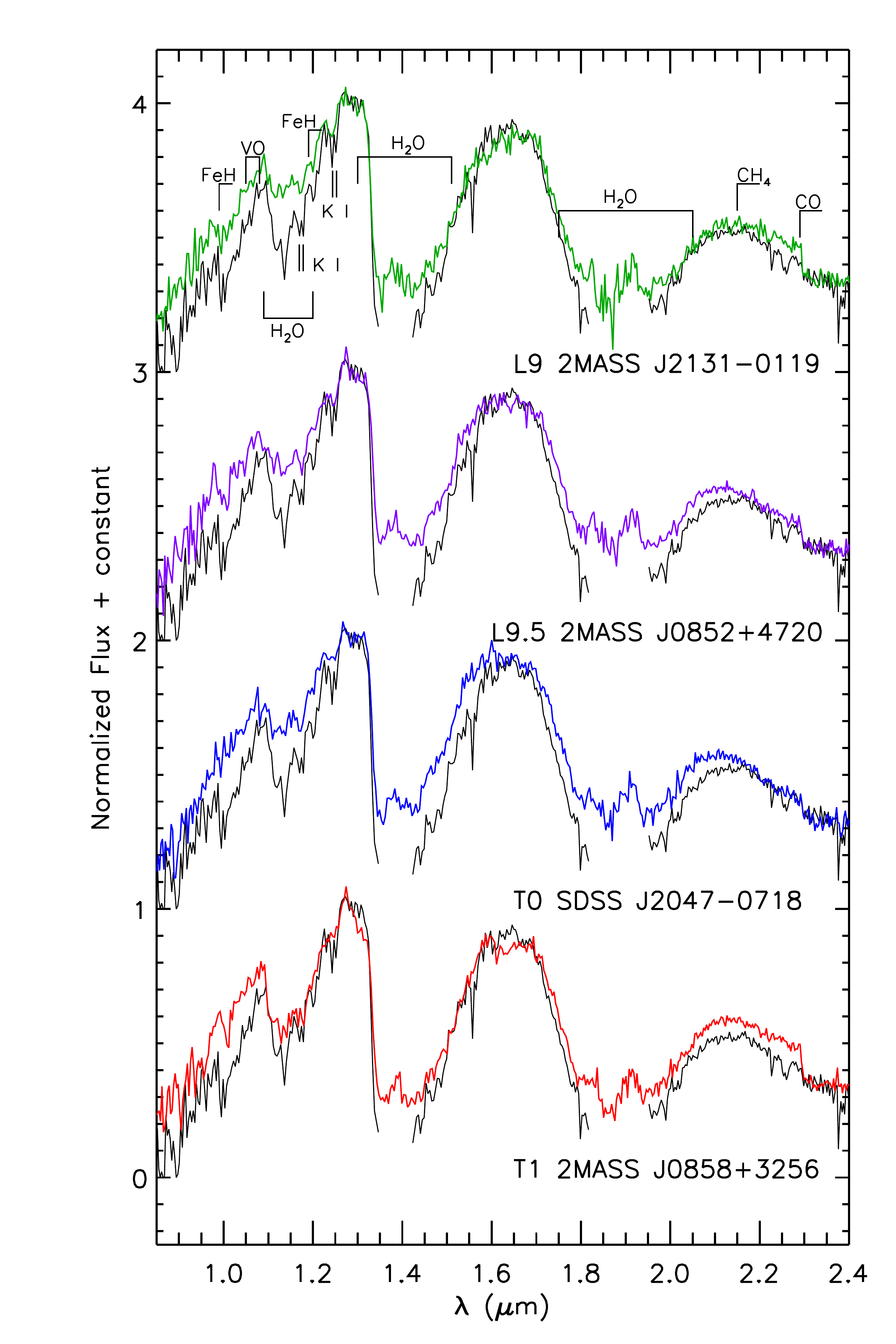}
\caption{\footnotesize Our FIRE spectrum of \target\ (black) compared to SpeX Prism Library spectra of L9--T1 dwarfs. The FIRE spectrum has been smoothed to the resolution of the SpeX spectra. Comparison spectra are: L9 (2MASS J21315444$-$0119374; \citealp{chiu06}); L9.5 (2MASS J08523490+4720359; \citealp{burgasser10}); T0 (SDSS J204749.61$-$071818.3; \citealp{burgasser10}); T1 (2MASS J08583467+3256275; \citealp{burgasser10}). Spectra are normalized to the average flux in the 1.25--1.30\micron\ region.}
\label{fig:spt}
\end{figure}%

In Figure~\ref{fig:spt} we compare our FIRE spectrum of \target\ to the closest matching spectra of each type of L9--T1 dwarf from the SpeX Prism Library. From this comparison, we see that \target\ has more FeH and \ion{K}{1} absorption than all of the comparison objects and very little CO absorption at $\sim$2.3\micron. We also see that \target\ has deeper water bands. 

All of these features are characteristic of an old, metal-poor ultracool dwarf \citep{burgasser08a}. Low-metallicity brown dwarfs typically have increased FeH absorption and weak signatures of CO. These metal poor objects are also old with high surface gravities which leads to greater line strengths of the pressure-sensitive alkali species, namely \ion{K}{1}. As \cite{burgasser08a} details, however, these characteristics cannot explain all of the features in the spectra of subdwarfs, particularly the strong H$_2$O bands. For that, thin and/or large-grain condensate clouds are needed to reduce the contrast between the $J$-band peak and 1.4\micron\ water band.

\begin{figure}[t!]
\centering
\includegraphics[width=1\linewidth]{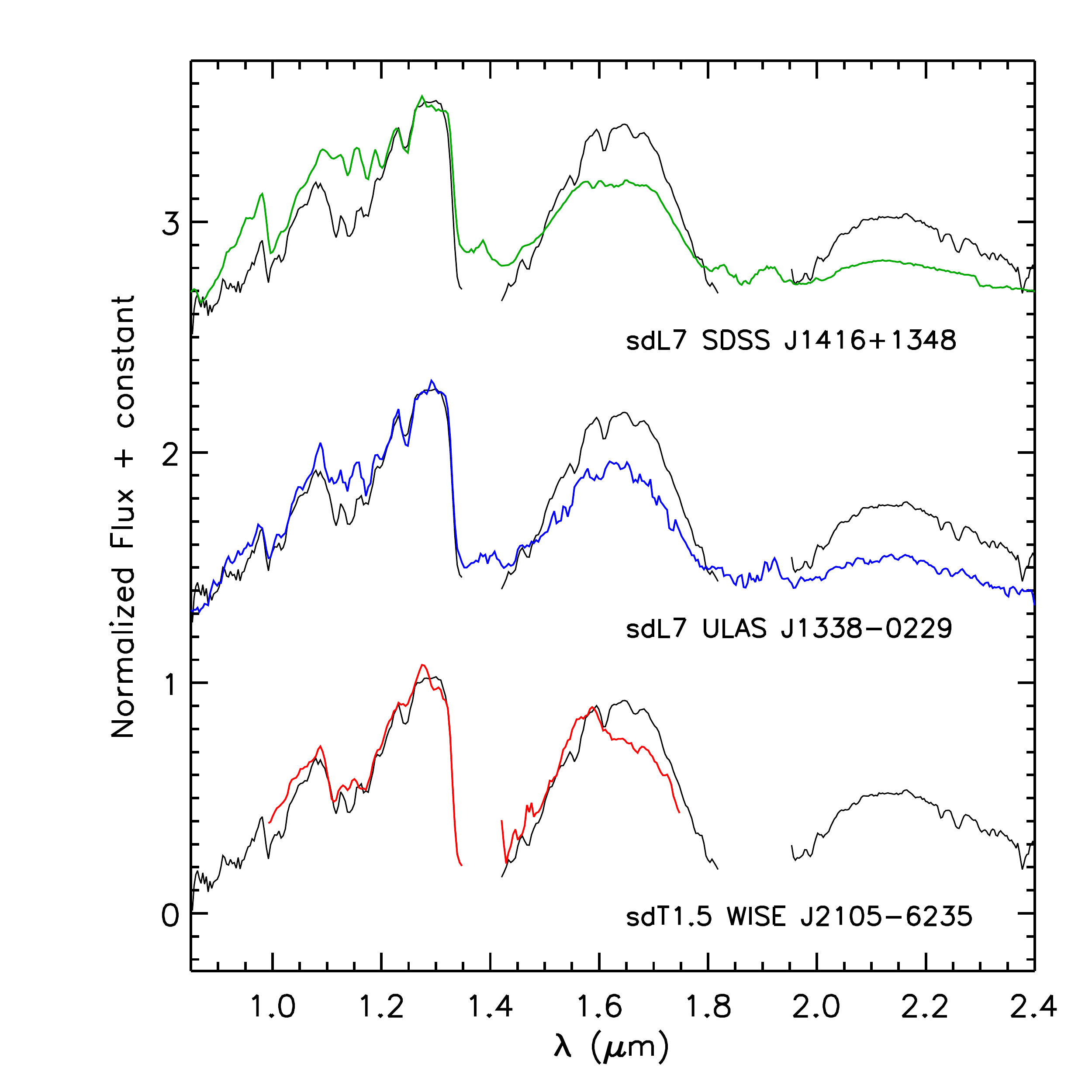}
\caption{\footnotesize Our FIRE spectrum of \target\ (black) compared to the SpeX Prism spectrum of SDSS J141624.08+134826.7 (sdL7; \citealp{schmidt10}), the FIRE Prism spectrum of ULAS J133836.97--022910.7 (sdL7; \citealp{zhang17a}), and the Gemini Flamingos-2 spectrum of WISE J210529.08$-$623558.7 (sdT1.5; \citealp{luhman14}). All spectra have been smoothed to the resolution of the SpeX spectrum for a more direct comparison. Spectra are normalized to the average flux in the 1.25--1.30\micron\ region.}
\label{fig:sdcomp}
\end{figure}

Although these are typical characteristics of low-metallicity L dwarfs, we assume they are also the features of late-L/early-T subdwarfs. One defining characteristic of subdwarfs that we cannot match is the blue $J-K$ color --- our NIR spectrum does not show enhanced collision-induced H$_2$ absorption (CIA H$_2$) in the $H$- and $K$-bands. If \target\ had enhanced CIA H$_2$ absorption, the $H$- and $K$-bands would also appear flatter as is seen in other subdwarfs (Figure~\ref{fig:sdcomp}). Since this is not the case we conclude that this object truly does not have a blue $J-K$ color and the redder $J-K$ slope is not a consequence of instrumental effects or an improperly calibrated spectrum. The derived color can be checked via future follow-up either by obtaining another higher signal-to-noise spectrum or by obtaining $J$ and $K$ photometry.

We have also compared our spectrum of \target\ to two L7 subdwarfs (SDSS J141624.08+134826.7, \citealp{schmidt10}; ULAS J133836.97--022910.7, \citealp{zhang17a}) and a suspected T1.5 subdwarf (WISE J210529.08$-$623558.7, \citealp{luhman14}) in Figure~\ref{fig:sdcomp}. These are a few of the only late-L/early-T subdwarfs known to date. From Figure~\ref{fig:sdcomp}, we see that the strengths of the \ion{K}{1} and FeH features of \target\ match those of the sdL7 dwarf ULAS J1338--0229 as well as the lack of CO absorption. Both SDSS J1416+1348 and ULAS J1338--0229 have bluer $J-K$ colors and shallower H$_2$O bands but this could be a consequence of the different spectral types.  The depths of the H$_2$O bands of \target\ are more comparable to those of WISE J2105--6235 but the \ion{K}{1} and FeH features are stronger. The long-wavelength side of the $H$-band appears to be suppressed by CIA H$_2$ but the effects on the $K$-band are unknown since the F2 spectrum of WISE J2105--6235 only covers the $J$- and $H$-bands.

Although \target\ lacks the characteristic blue NIR color of a typical low-metallicity object, all other signs point to it being part of the old galactic population. Thus, we tentatively conclude that this is a T0$\pm$1 subdwarf.

\begin{figure*}[t!]
\centering
\includegraphics[width=1\linewidth]{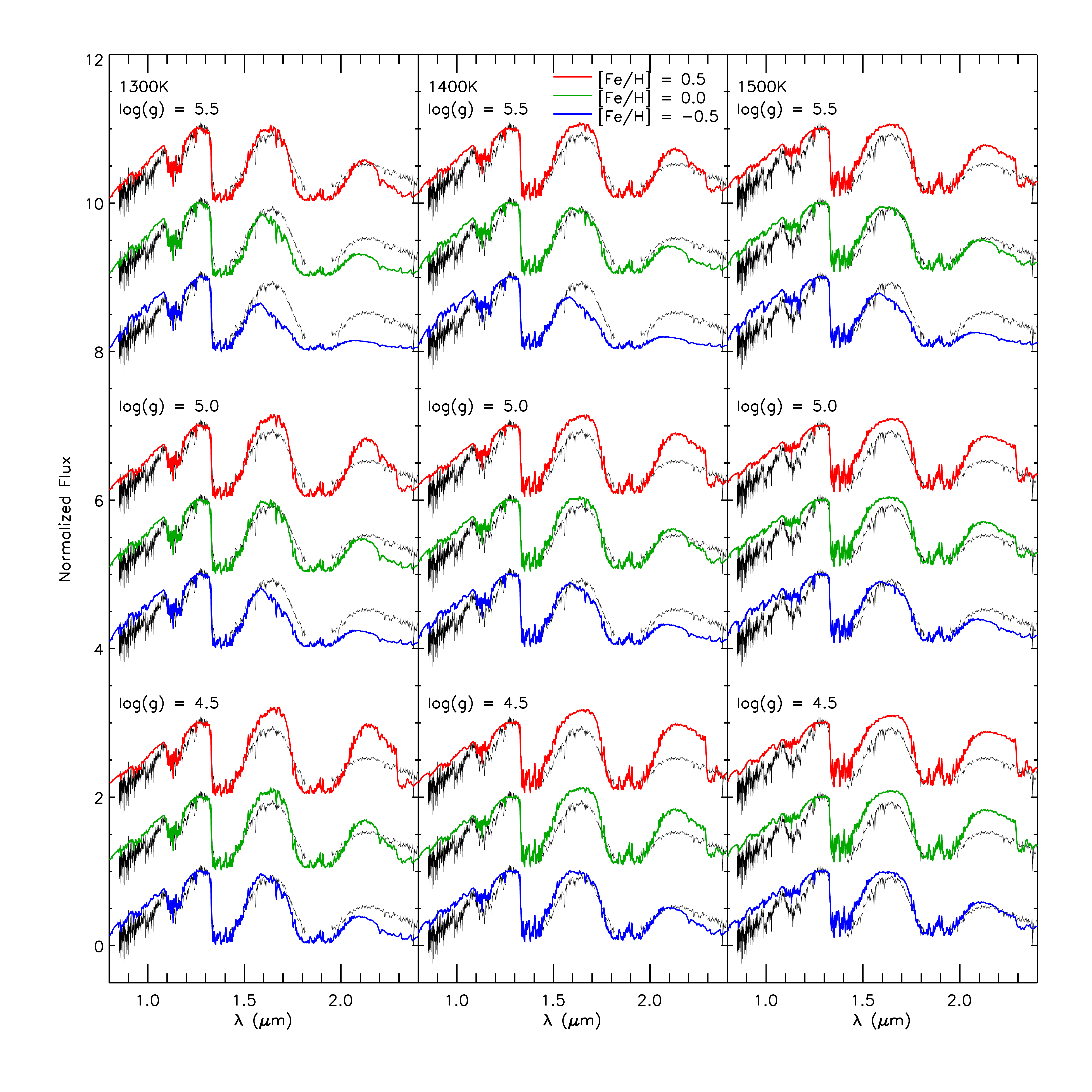}
\caption{\footnotesize The FIRE spectrum of \target\ (black) compared to theoretical spectra from \cite{burrows06}. The low-metallicity models (blue) and low-gravity models (lowest sets) seem to fit the spectrum of \target\ best, however, the fits are quite poor. Spectra are normalized to the average flux in the 1.25--1.30\micron\ region.}
\label{fig:models}
\end{figure*}

We have also compared our spectrum to theoretical spectra. The set of models that most comprehensively cover the temperature, gravity and metallicity parameters are the spectra from \cite{burrows06}. We have compared our FIRE spectrum of \target\ to model spectra that have \Teff=1300--1500K, log $g$=4.5--5.5 and metallicities between 0.3 and 3 times solar (Figure~\ref{fig:models}). We can see from Figure~\ref{fig:models} that when the models are compared to each other, the low-metallicity models have deeper FeH and \ion{K}{1} features and a lack of CO absorption compared to the higher metallicity models. Since this is similar to what we observe when we compare \target\ to other L and T dwarfs, we reason that this object likely also has a low metallicity, despite the poor model fits and apparent lack of enhanced CIA H$_2$ absorption. The lower surface gravity models appear to provide better comparisons visually but the fits are still quite poor. The fact that the best fits appear to have both a low surface gravity and a low metallicity is puzzling as we would expect a low-metallicity object to be older, and hence, have higher surface gravity. 

The overall best fitting model is the 1500K, log $g$=5.5 and 0.3 $\times$ solar metallicity model (which we will denote as 15-5.5-0.3X) with a reduced $\chi^2$ of 0.19. The best fitting models in the $J-$, $H-$, and $K-$ bands are the 15-5.5-0.3X (reduced $\chi^2$ = 0.17), 15-4.5-0.3X (reduced $\chi^2$ = 0.16) and 14-4.5-3X (reduced $\chi^2$ = 0.42), respectively. The inconsistency between the best-fit models and the poor visual similarity of these best-fit models with the spectrum of \target, does not allow us to make any definitive conclusions about the nature of this object.

\subsection{Unresolved Multiple?}

\begin{figure*}
\centering
\includegraphics[width=1\linewidth]{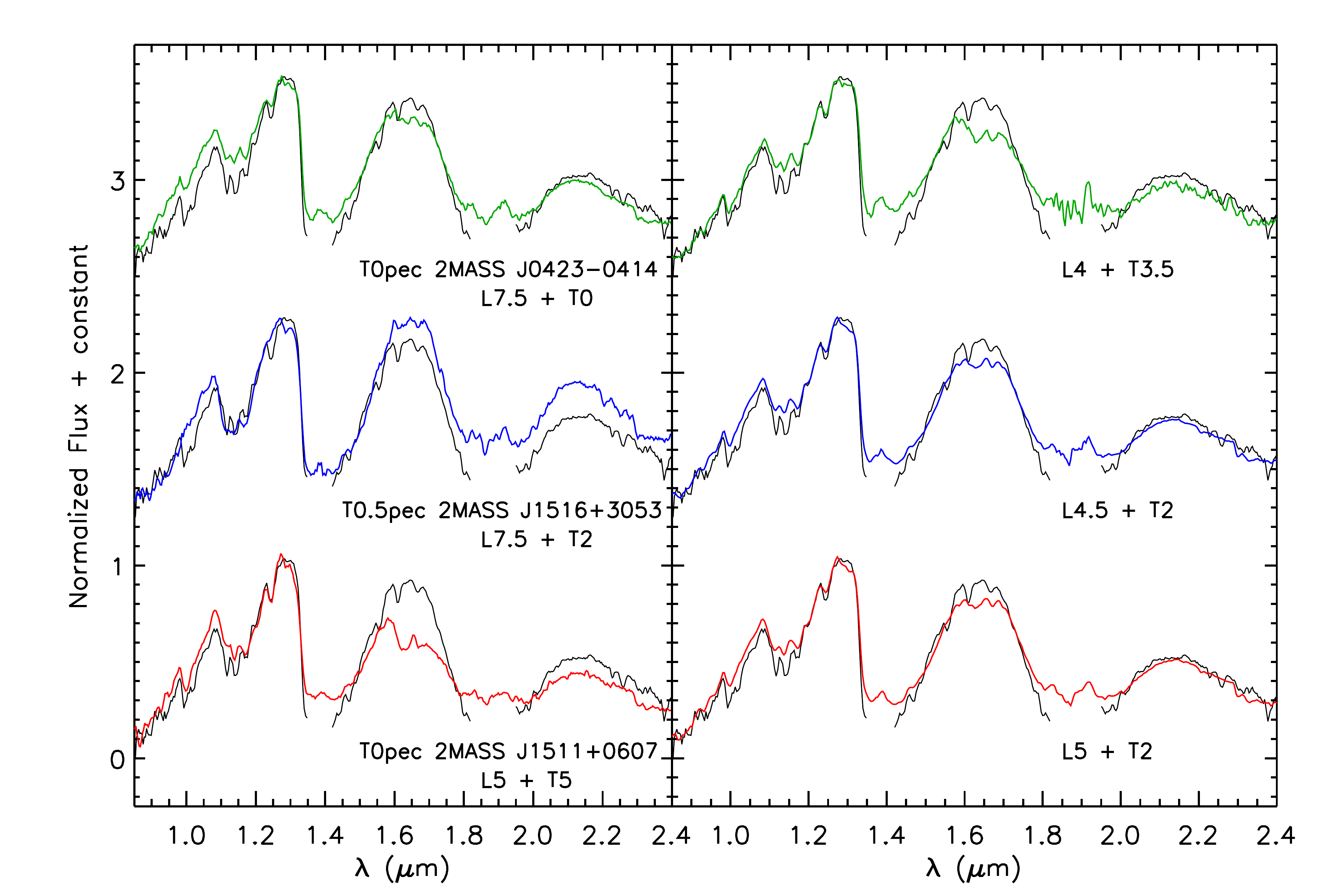}
\caption{\footnotesize Our FIRE spectrum of \target\ (black) compared to SpeX Prism peculiar spectra (left) and spectral binary templates (right). The FIRE spectrum has been smoothed to the resolution of the SpeX spectra. Comparison spectra for the peculiar objects are: T0pec (2MASS J04234858$-$0414035; \citealp{burgasser04}); T0.5pec (SDSS J151642.97+305344.5; \citealp{burgasser10}); T0pec (2MASS J15111466+0607431; \citealp{chiu06}). The spectra used in the spectral binary templates are: L4 (2MASS J04070752+1546456; \citealp{burgasser08b}) and T3.5 (SDSS J153417.05+161546.1; \citealp{chiu06}); L4.5 (2MASSI J0652307+471034; \citealp{burgasser10}) and T2 (SDSS J024749.90--163112.6; \citealp{chiu06}); L5 (2MASSW J1239272+551537; \citealp{burgasser10}) and T2 (SDSS J024749.90--163112.6; \citealp{chiu06}). Spectra are normalized to the average flux in the 1.25--1.30\micron\ region.}
\label{fig:sptpec}
\end{figure*}

We also compared the spectrum of \target\ to objects that were classified as peculiar in the SpeX Prism Library and to templates of spectral binaries. We created our templates by normalizing all single L and T dwarfs in the SpeX Prism Library to the flux in the 1.25--1.30\micron\ region, scaling them to their spectral-type dependent absolute magnitudes given in \cite{filippazzo15}, and summing the pairs of resulting spectra. The best matches are shown in Figure~\ref{fig:sptpec}, however, none of the spectra can reproduce all of the features in the spectrum of \target, in particular the slopes in the $H$-band. Interestingly, the best matching peculiar spectra in the $J$-band are 2MASS J04234858$-$0414035 \citep{burgasser04}, 2MASS J15111466+0607431 \citep{chiu06} and SDSS J151642.97+305344.5 \citep{burgasser10}, all of which are confirmed or suspected binaries \citep{burgasser06,burgasser10,bardalez15}. The L7.5 + T0 spectrum of 2M J0423$-$0414 fits the $H$- and $K$-band quite well but matches the $J$-band only marginally better than the single spectra of Figure~\ref{fig:spt}. The L7.5 + T2 spectrum of SDSS J1516+3053 matches the depth of the water absorption bands but fails to match the FeH and \ion{K}{1} absorption features and the overall slope of \target. The L5 + T5 spectrum of 2M J1511+0607 fits the $J$-band of \target\ extremely well but fails to match the $H$- and $K$-bands, mostly due to the contribution of the T5 in 2M J1511+0607. 

From these comparisons, we see that as the fit in the $J$-band improves with combinations of earlier L dwarfs and later T dwarfs, the fit in the $H$- and $K$-bands becomes worse. This, along with the apparent lack of CH$_4$ absorption disfavors the possibility that this object is a spectral binary. The redder $J-K$ slope of \target\ compared to subdwarfs and lack of CIA by H$_2$, however, potentially supports the binary explanation.

The best fits from the spectral binary template fitting seem to produce better matches (right panel of Figure~\ref{fig:sptpec}), however, the templates still cannot match the strength of the \ion{K}{1} and FeH features. Quantitative measurements show these templates provide no, or no significant, improvement over the fit to the sdT1.5 dwarf WISE 2105--6235 (reduced $\chi^2$ = 0.30, 0.17 and 0.17, respectively, versus 0.19 for WISE 2105--6235). We also note that at least one spectrum in each of the templates is itself a suspected binary or other peculiar object: the T3.5 dwarf SDSS J153417.05+161546.1 \citep{chiu06} is a T1.5+T5.5 binary \citep{liu06}; the L4.5 dwarf 2MASSI J0652307+471034 \citep{burgasser10} is potentially young \citep{cruz07}; the T2 dwarf SDSS J024749.90--163112.6 \citep{chiu06} is a candidate T0+T7 binary \citep{burgasser10}; and the L5 dwarf 2MASSW J1239272+551537 \citep{burgasser10} is an L5+L6 binary \citep{radigan13}. Because one spectrum in each template is itself a suspected binary, the spectra are in fact triple templates. Since the scenario of a young or thin-disk triple system moving with such large proper motion is unlikely, we also tentatively rule out this possibility. However, we cannot altogether dismiss the possibility of unresolved multiplicity.

\subsection{Kinematics} \label{subsec:kin}

Using the absolute magnitude vs. spectral type polynomial from \cite{faherty16} for a T0 $\pm$ 1 dwarf, we estimate that \target\ has an absolute $W2$ magnitude of $\sim$11.8 mag, giving a photometric distance of 37 pc. The proper motion of this object at the photometric distance corresponds to a tangential velocity of $\sim$170 \kms. The relations of \cite{faherty16}, however, were calibrated using normal field objects and we expect the absolute $W2$ magnitudes of late-L/early-T subdwarfs to be slightly fainter than those of normal objects of the same spectral type as that is the behavior seen in higher mass brown dwarfs and low-mass stars. This would mean our distance and velocity values are slightly overestimated. Using the absolute magnitude vs. spectral type relations from Gonzales et al.\ (2017; in prep) for subdwarfs up to L5 and doing a naive extension to T0 spectral types, \target\ should have an absolute magnitude of $\sim$12.3 mag, giving a photometric distance of 29 pc. The proper motion then corresponds to $\sim$150 \kms; a much smaller value. 

WISE has an angular resolution of 6\farcs1 in W1 and 6\farcs4 in W2. If \target\ is a multiple object, the components would have a separation of less than 226 $\pm$ 26 AU or 177 $\pm$ 26 AU, using the two distance estimates 37pc and 29pc respectively, to be unresolved with WISE.

\begin{table}
\centering
\begin{tabularx}\textwidth{cc}
\caption{Inferred Properties of \target} \label{tab:target_inf} \\
\hline \hline \\[-4ex]

\endhead

Spectral Type & sdT0 $\pm$ 1 \\
\hline \\[-4ex]
From \cite{faherty16} & \\
\hline \\[-4ex]
M$_{W2}$ & 11.8 $\pm$ 0.5 mag \\
Photometric Distance & 36.8 $\pm$ 4.2 pc \\
Tangential Velocity & 173 $\pm$ 22 km\,s$^{-1}$\\
\hline \\[-4ex]
From Gonzales et al.\ (2017; in prep) & \\
\hline \\[-4ex]
M$_{W2}$ & 12.3 $\pm$ 0.5 mag \\
Photometric Distance & 28.9 $\pm$ 4.2 pc \\
Tangential Velocity & 149 $\pm$ 22 km\,s$^{-1}$ \\
\hline \\[-4ex]
\end{tabularx}
\centering {\footnotesize Uncertainties represent the values for the range of possible spectral types --- L9--T1.}
\end{table}

\section{Halo/Thick-Disk Brown Dwarfs} \label{sec:halo}

From studies of stars, tangential velocities of objects in the thick-disk have been found to be $\sim$85--180~\kms\ (e.g. \citealp{fuhrmann98,feltzing03,soubiran03}) and $\sim$200--300~\kms\ in the halo (e.g. \citealp{chiba00,mace13,schilbach09}). With a tangential velocity of $\sim$170~\kms, \target\ lies right on the cusp between these populations. Based on the membership probability distributions calculated by \cite{dupuy12}, \target\ has a $>$90\% probability of being part of the thick-disk and $<$10\% probability of being a member of either the halo or thin-disk populations. Using the tangential velocity calculated using the relation from Gonzales et al.\ (2017; in prep), \target\ has a membership probability of 50--90\% for the thick-disk, $\leq$10\% for the thin-disk, and $<$10\% for the halo. If \target\ were an unresolved binary, it would be a factor of $\sim$30--40\% further away, and so will have a proportionately higher tangential velocity: putting it further into the thick-disk kinematics.

The comparison of \target\ with solar-metallicity objects and theoretical spectra with [Fe/H] values between 0.0 and $-0.5$ indicates a low metallicity, making it consistent with either the thick-disk or halo populations \citep{reddy06,lepine07,zhang17b}. However, this is only tentative as the model fits are quite poor.

Only a handful of T dwarfs have been confirmed to be a part of the thick-disk or halo populations, most of which are late T dwarfs --- WISE J200520.38+542433.9 (T8; v$_{tan}\approx$ 110 \kms; \citealp{mace13}), WISE J001354.39+063448.2 (T8; v$_{tan}$\footnote{Assuming a photometric distance using mid-infrared magnitudes as we do in this work.} $\approx$ 107--190 \kms) and WISE J083337.83+005214.2 (T9; v$_{tan}$\footnotemark[3] $\approx$ 126--231 \kms; \citealp{pinfield14}), and ULAS J131610.28+075553.0 (T6.5; v$_{tan}\approx$ 240--340 \kms; \citealp{burningham14}). The latest L-type brown dwarfs that are part of the thick-disk or halo populations are L7 dwarfs (e.g. \citealp{burgasser03,zhang17a}). There has, so far, been a lack of discoveries of low-metallicity late-L and early-T brown dwarfs. Only one early-T dwarf, WISE J210529.08$-$623558.7 (T1.5; v$_{tan}$ = 176 $\pm$ 25 \kms; \citep{luhman14}, has been found to potentially have kinematics and a metallicity consistent with these populations. The discovery of \target\ potentially brings the number of objects in this observational gap up to two. 
 
More detailed metallicity measurements and kinematics are needed to confidently determine to which population this object belongs: halo or thick-disk.
 
\section{Conclusions} \label{sec:concl}

We have determined that our candidate object, WISE J071121.36--573634.2, is potentially a sdT0 $\pm$ 1 dwarf. The estimated absolute $W2$ magnitude of this object puts it at $\sim$37 pc with a tangential velocity of $\sim$170~\kms, or $\sim$29pc with a tangential velocity of $\sim$150~\kms. The properties of \target\ are generally consistent with those of the thick-disk population. Using either of the distance estimates, this object is relatively close to the sun and so obtaining a parallax measurement should be relatively easy. Provided this object is indeed an sdT0 dwarf, the discovery of \target\ would only be the second low-metallicity late-L/early-T dwarf discovered to date. Further spectroscopic observations will allow us to be able to definitively determine the metallicity, thus the membership, of this object. New observations will also allow us to rule out (or confirm) multiplicity.

\acknowledgements
 
We thank the anonymous referees for their very constructive comments that helped improve this paper. KK acknowledges support via the IPAC Visiting Graduate Student Program. This publication makes use of data products from the Wide-field Infrared Survey Explorer, which is a joint project of the University of California, Los Angeles, and the Jet Propulsion Laboratory/California Institute of Technology, funded by the National Aeronautics and Space Administration. 

\facilities{Magellan:Baade (FIRE)}
\software{IDL}

\bibliography{bibliography.bib}
\bibliographystyle{apj}

\end{document}